# Near-Infrared and Visible-range Optoelectronics in 2D Hybrid Perovskite/Transition Metal Dichalcogenide Heterostructures

Abin Varghese,[1,2,3,4] Yuefeng Yin,[1,4] * Mingchao Wang,[1,5] Saurabh Lodha,[2] Nikhil V. Medhekar[1,4] *

Email: yuefeng.yin@monash.edu, nikhil.medhekar@monash.edu

[1] Department of Materials Science and Engineering, Monash University, Clayton, Victoria 3800, Australia

[2] Department of Electrical Engineering, Indian Institute of Technology Bombay, Mumbai 400076, India

[3] IITB-Monash Research Academy, IIT Bombay, Mumbai 400076, India

[4] ARC Centre of Excellence in Future Low-Energy Electronics Technologies, Monash University, Clayton, Victoria 3800, Australia

[5] Centre for Theoretical and Computational Molecular Science, Australian Institute for Bioengineering and Nanotechnology, The University of Queensland, St Lucia, QLD 4072, Australia

**Abstract**

The application of ultrathin two-dimensional (2D) perovskites in near-infrared and visible-range optoelectronics has been limited owing to their inherent wide bandgaps, large excitonic binding energies and low optical absorption at higher wavelengths. Here, we show that by tailoring interfacial band alignments via conjugation with low-dimensional materials like monolayer transition metal dichalcogenides (TMD), the functionalities of 2D perovskites can be extended to diverse, visible-range photophysical applications. Based on the choice of individual constituents in the 2D perovskite/TMD heterostructures, our first principles calculations demonstrate widely tunable type-II band gaps, carrier effective masses and band offsets to enable an effective separation of photogenerated excitons for enhanced photodetection and photovoltaic applications. In addition, we show the possibilities of achieving a type-I band alignment for recombination based light emitters as well as a type-III configuration for tunnelling devices. Further, we evaluate the effect of strain on the electronic properties of the heterostructures to show a significant strain tolerance, making them prospective candidates in flexible photosensors.

# Introduction

Two-dimensional (2D) perovskites are rapidly emerging as useful materials for optoelectronic applications because of their direct bandgaps,[1,2] high photoluminescence quantum yield,[1] broadband light emission[3] and improved stability with respect to their bulk counterparts.[4,5] They are represented by the structural formula $A'_2A_{n-1}M_nX_{3n+1}$, where A' and A are organic cations, M is transition metal (commonly Pb), and X is halogen.[1,2] In a single unit of the 2D perovskite, the 2D metal-halide octahedra is sandwiched within bulky organic spacer cations, A', forming a hybrid organic-inorganic configuration. *n* determines the number of such metal halide units stacked vertically, and A represents the smaller organic cation.[1] Adjacent such units are held by weak van der Waals (vdW) interactions, rendering the possibility of micromechanical exfoliation to produce atomically thin monolayers similar to graphene-like 2D materials.[6]

In contrast to other 2D vdW systems, the dielectric confinement due to bulky organic groups gives rise to large bandgap values in monolayer (*n*=1) 2D perovskites. For example, photoluminescence spectrum of the well-explored $BA_2PbBr_4$ (BA is butylammonium, $C_4H_9NH_3$) yields an optical bandgap 3 eV,[1] while in $BA_2PbI_4$ and $(C_6H_5CH_2NH_3)_2PbCl_4$ the bandgaps are reported to be 2.4 eV[2] and 3.6 eV[7], respectively. Such wide bandgaps lead to a poor absorption of light in the near-infrared and visible regions of the electromagnetic spectrum and limit the applications of 2D perovskites in solar cells and visible-range light emission or detection. The out-of-plane quantum well-like configuration also leads to large exciton binding energies in the range of 200-500 meV.[8,9] This makes the separation of photogenerated excitons into free carriers less feasible. However, to achieve higher photovoltaic efficiencies, an effective bandgap near the Shockley-Queisser range (1.3–1.5 eV) as well as an easy dissociation of excitons into free carriers are required.

A possible solution to overcome these limitations is to design heterostructures of 2D perovskites with suitable materials that demonstrate excellent optical properties and can lead to desired interface properties like faster exciton separation.[10,11] Transition metal dichalcogenides (TMDs) are well-studied van der Waals materials with exceptional electronic and optical properties, sizeable bandgaps, and high tensile strength.[12] This makes them suitable candidates to conjugate with 2D perovskites to form van der Waals heterostructures without the need of a coherent epitaxial growth. The absence of dangling bonds in both TMDs and 2D perovskites can result in clean heterointerfaces. Such graphene/TMD/III–VI metal chalcogenide/hBN based vdW heterostructures have led to the emergence of unprecedented physical phenomena ranging from electrical field-effect[13] and tunnelling devices[14] to light detectors, harvesters,[15] and emitters[16] to twistronics[17] based novel systems which are all governed by the hetero-interface.[11] Therefore, van der Waals heterostructures formed using 2D perovskites and TMDs need be explored to understand the nature of the interface and emergence of new functionalities.

From a device aspect, most experimental reports have focused on the improved photodetection in these heterostructures, whereas photovoltaic performance and light emission are not well investigated.[18,19] Recent studies of 2D perovskite/TMD heterostructures have shown interlayer exciton peaks,[20] excitonic energy transfer[21] as well as significant photoluminescence quenching at the interface,[18] indicating a possibility of charge transfer and strong interlayer coupling between these structurally dissimilar materials. While doped-graphene and phosphorene have been proposed for forming heterojunctions with 2D perovskites, they are limited by the absence of bandgap[22] and poor air stability and tunability,[23] respectively. However, for specific predefined applications, the rational choice of individual TMD and 2D perovskite constituents of the heterostructure should be based on their interfacial electronic bandstructure.[24] A possible diverse range of band alignments at the interface can enable various applications like photodetection, photovoltaic operation or light emission owing to the strong light-matter interaction in individual materials.

In this work using first principles calculations, we illustrate a number of heterostructures of 2D perovskites with transition metal dichalcogenides that can result in new device functionalities and extend the optical bandwidth of the 2D perovskites for improved optoelectronic performance. In particular, the emergence of a type-II band alignment at the interface of $BA_2PbBr_4$/$MoS_2$ heterojunction with a lower effective bandgap in the near-infrared region along with the possibility of enhanced interlayer separation of photogenerated charge carriers can result in improved photodetection and photovoltaic performance. In a heterostructure of $BA_2PbBr_4$ with $WSe_2$, the possibility of a type-I band alignment could enable carrier recombination for light-emitting devices. We discuss the chalcogen-dependent properties of various TMD heterojunctions with the homologous series of 2D perovskites $BA_2PbBr_4$, $BA_2PbI_4$ and $BA_2PbCl_4$ and show that tunnelling-based (type-III) devices are also possible by the apt choice of constituents. Finally, we show that 2D perovskite/TMD heterostructures can withstand considerable tensile strains without significant degradation of the electronic properties potentially enabling their applications on stretchable substrates for flexible devices.

## Results & Discussion

### Structure analysis:

We explore the interface of various Group-6 TMDs ($MoS_2$, $WS_2$, $MoSe_2$, and $WSe_2$) with 2D perovskites, $BA_2PbCl_4$, $BA_2PbBr_4$ and $BA_2PbI_4$ through first principles-based density functional theory (DFT) calculations. $BA_2PbX_4$ (X=Cl/Br/I) is a $n = 1$ 2D perovskite made up of the inorganic $[PbX_4]^{2-}$ octahedra incorporated between the large BA cations as shown in the schematic in Figure 1a. The hybrid 2D perovskites and group-6 TMDs discussed in this work have orthorhombic and hexagonal unit cells, respectively. The optimized lattice parameters for all perovskites and monolayer TMDs are provided in Supporting Information S1. Additionally, we also compare the lattice parameters and bandgap values using different van der Waals corrected methods (DFT-D3, DFT-D2 and optB86vdW) in S1.

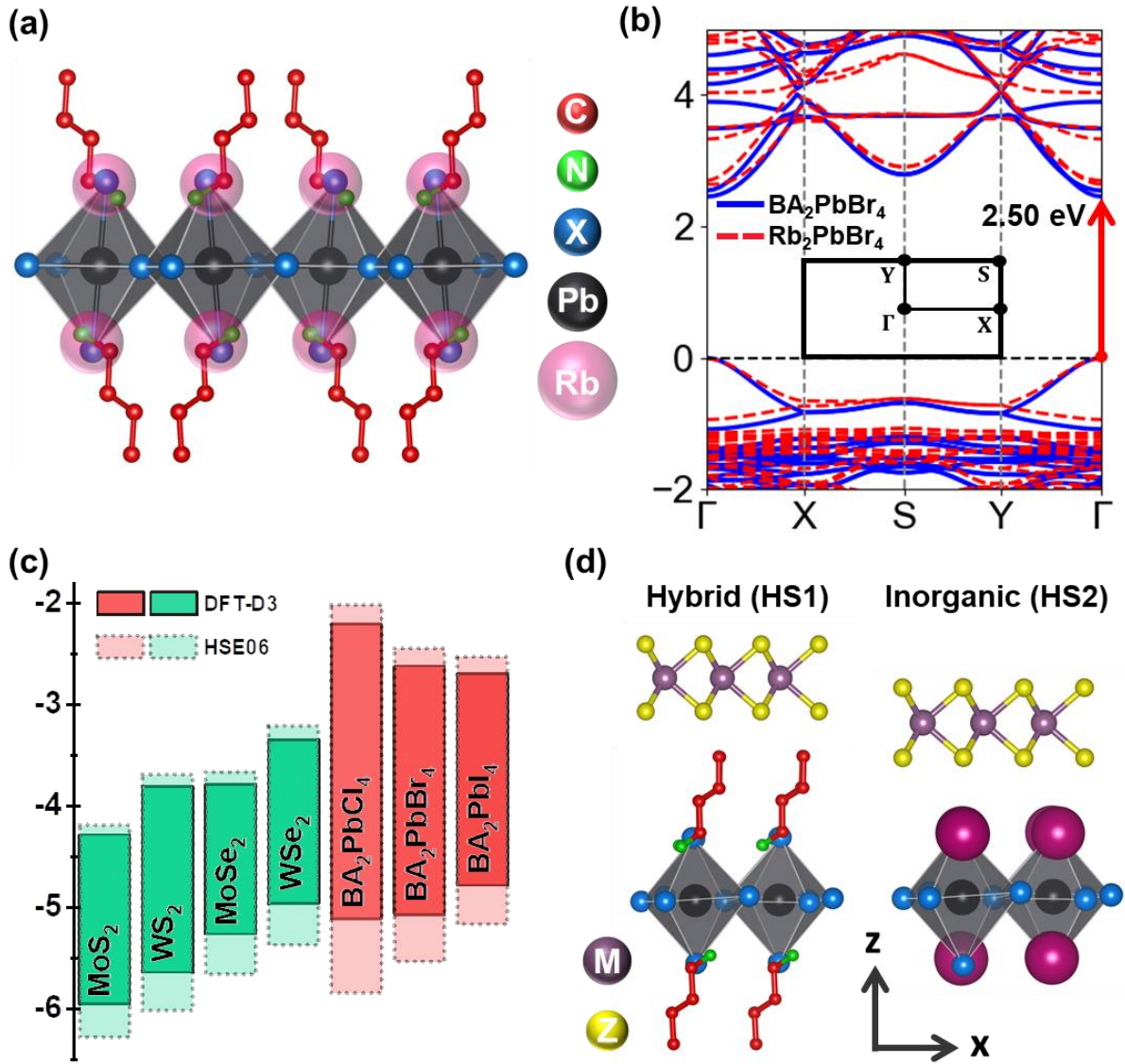

Figure 1: (a) Schematic representation of *n*=1 hybrid organic-inorganic 2D perovskite-BA$_2$PbX$_4$. X is halogen, H atoms are not shown for clarity. Rb atoms are represented by transparent spheres to substitute organic chains to form all-inorganic 2D perovskite Rb$_2$PbX$_4$. (b) Comparison of the bandstructures of hybrid organic-inorganic BA$_2$PbBr$_4$ (solid blue) and fully inorganic Rb$_2$PbBr$_4$ (dashed red) showing excellent agreement in band features at the band edges. Fermi level is set at the valence band maximum. (c) Pre-contact band alignments of various TMDs and 2D perovskites employed in this study calculated using van der Waals corrected DFT-D3 and hybrid HSE06 methods. (d) Two distinct configurations involving 2D perovskites with TMDs, namely, HS1 being the hybrid organic-inorganic heterostructure, and HS2 being the fully inorganic heterostructure with Rb replacing the organic moiety.

**<u>Electronic properties:</u>**

Figure 1b shows the bandstructure of single layer $BA_2PbBr_4$ as a representative of the 2D perovskites considered here. As also illustrated in the Supporting Information S2, the bandstructures of all 2D perovskites depict direct bandgaps with the conduction band minimum (CBM) and valence band maximum (VBM) occurring at **Γ** of the orthorhombic Brillouin zones (BZ). The direct bandgap values calculated using both generalised gradient approximation (GGA) as well as Heyd-Scuseria-Ernzerhof (HSE06) formalism are found to be in the near-ultraviolet region and even higher. As an example, incorporating the van der Waals corrected DFT-D3, the DFT-GGA and HSE06 bandgaps of $BA_2PbBr_4$ are 2.50 eV and 3.37 eV, respectively, while the experimental value is reported to be around 3 eV.[1,8,25] On the other hand, the bandgaps in monolayer TMDs (Supporting Information S3) are in the range of 1.5 eV–1.9 eV (DFT-GGA) and 1.9 eV–2.4 eV (HSE06). For example, our calculations show the DFT-GGA bandgap in monolayer $MoS_2$ to be 1.65 eV (HSE06 yields 2.23 eV) and the experimentally obtained optical bandgap using photoluminescence (PL) is near 1.9 eV.[26] Further, the pre-contact band alignments of the TMDs- $MoS_2$, $WS_2$, $MoSe_2$, and $WSe_2$, as well as the 2D perovskites employed in this study, calculated using the respective work functions, are shown in Figure 1c. In general, the bandgap values obtained by using HSE06 formalism are closer to the experimental values, consistent with several reports.[24,27]

We model 2D perovskite/TMD heterostructures using an orthorhombic supercell in order to ensure a good lattice matching between the 2D perovskite and TMD layers. Each supercell consists of a 2×2 unit cell of the perovskite and 5×3 rectangular unit cell of the TMD layer, aligned in such a way to ensure a minimum possible strain and to enable a uniform comparison of electronic properties across different heterostructures. Both 2D perovskites and TMDs demonstrate significant strain tolerance, hence in the resulting heterostructure supercell the strain is split equally between the two layers.[28,29] In most configurations (see Supporting Information S4), the in-plane strain values are low, which is not expected to significantly impact the electronic characteristics.

A closer look at the electronic density of states of $BA_2PbX_4$ in the Supporting Information S2 shows that the conduction and valence band edges of the hybrid 2D perovskites are constituted by the p-orbitals of Pb and halogen, respectively, and the states contributed by the organic species (C, H, N) are far away from the Fermi level.[25,30] Therefore, we test the feasibility of an all-inorganic 2D perovskite structure by substituting the organic BA groups in the hybrid system by a single large inorganic monopositive cation like Rb (shown as pink spheres in Figure 1a) and retaining the lattice parameters of the corresponding hybrid 2D perovskite.[30,31] The calculated bandstructure of the all-inorganic $Rb_2PbBr_4$ shows a good resemblance to that of the hybrid perovskite, depicted using red dashed lines in Figure 1b. Furthermore, our results show that substituting the organic layers on one side of the PbX octahedra (Figure 1a) results in a mixed organic-Rb structure ($BARbPbBr_4$) which also demonstrates an excellent resemblance to the bandstructure characteristics in Figure 1c (Supporting Information, S5). Such substitutions can significantly reduce the computational cost involved in investigating the 2D perovskite/TMD supercells. Further studies are required to explore the experimental feasibility and stability of such Ruddlesden-Popper or Dion-Jacobson based all-inorganic 2D perovskites.

Next, to study the interaction between the 2D perovskite and TMD in the heterostructure, we first employ the hybrid organic-inorganic structure for the 2D perovskite (HS1). We also consider a second heterostructure made up of the all-inorganic Rb-substituted 2D perovskite structure (HS2). We compare the nature of bands and bandgap values for both configurations by using the illustrative example of MoS$_2$ as the TMD layer.

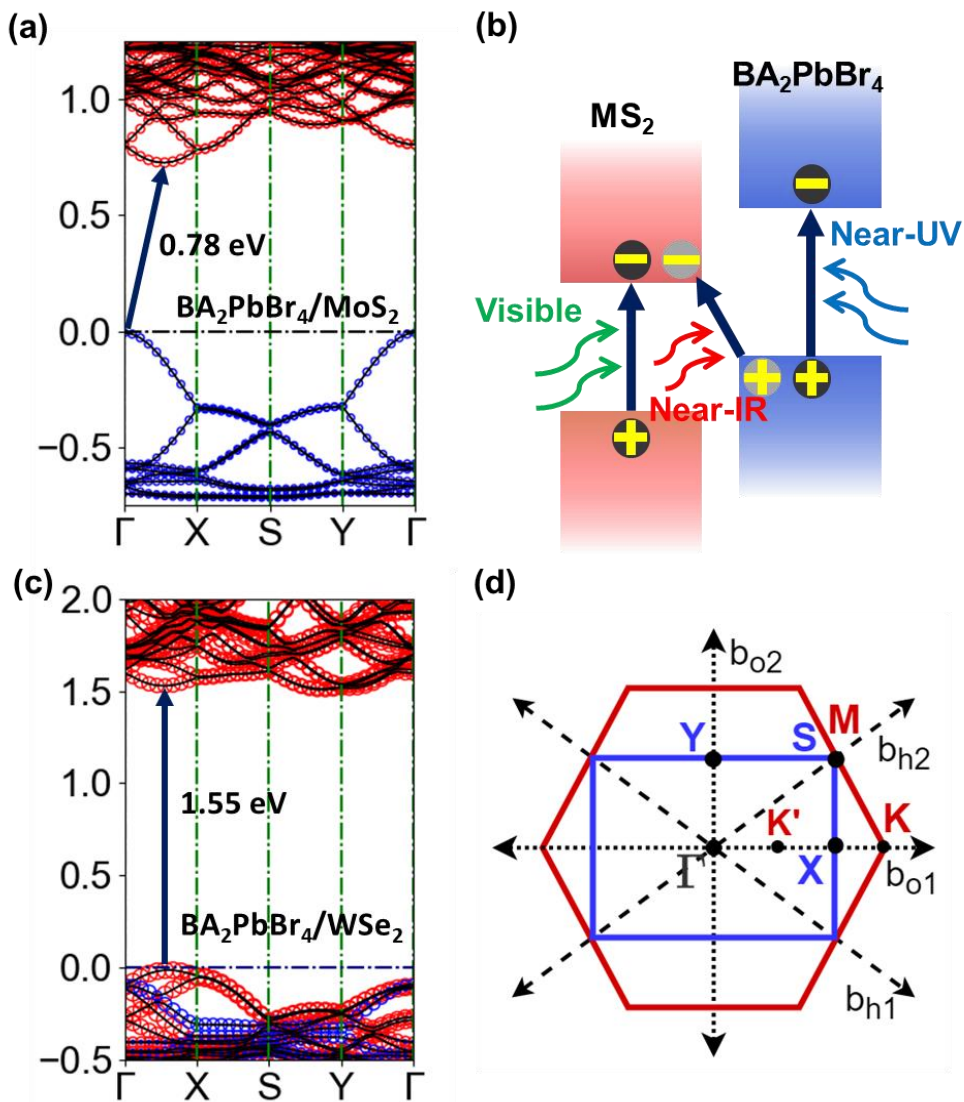

Figure 2: Bandstructure of 2D perovskite/TMD heterojunction. (a) Bandstructure of BA$_2$PbBr$_4$/MoS$_2$ heterostructure showing an indirect bandgap. The contributions from MoS$_2$ and BA$_2$PbBr$_4$ are marked using red and blue circles, respectively. (b) Schematic showing the various photogeneration possibilities in the heterostructure of BA$_2$PbBr$_4$ with sulphide based TMDs (MoS$_2$ and WS$_2$). (c) Bandstructure of BA$_2$PbBr$_4$/WSe$_2$ showing a nearly direct bandgap (difference between direct and indirect gaps is very low) along $\mathbf{\Gamma - X}$. Here too, the contribution from WSe$_2$ is in red and that of the perovskite is in blue. (d) Hexagonal and rectangular Brillouin zones of the respective unit cells of the TMD and 2D perovskite (with high symmetry points marked). b$_{o1}$, b$_{o2}$ represent the orthogonal axes of the orthorhombic cell and b$_{h1}$ and b$_{h2}$ represent those of the hexagonal cell. Due to the folding of the hexagonal Brillouin zone of TMDs onto the rectangular Brillouin zone of 2D perovskites, the high symmetry point $\mathbf{K}$ of the TMDs appears along $\mathbf{\Gamma - X}$ direction of the rectangular Brillouin zone of 2D perovskites.

The bandstructure of the hybrid organic-inorganic $BA_2PbBr_4/MoS_2$ heterostructure (HS1) is shown in Figure 2a, where the DFT-GGA calculated interface bandgap of 0.78 eV is significantly lower than the gaps of the individual 2D constituents of the heterostructure (1.9 eV for $MoS_2$, 3.0 eV for $BA_2PbBr_4$). From the bandstructure of the all-inorganic $Rb_2PbBr_4/MoS_2$ (HS2) shown in Supporting Information S6, the nature of the bands is identical to that in Figure 2a and the calculated gap value of 0.73 eV is in good agreement with HS1. Further, to elucidate the nature of interaction between the two materials we also examine a mixed organic-Rb substituted structure (HS3, Supporting Information S5) in which the organic groups away from the herointerface are substituted by Rb. The similarity in band features in the three cases proves that the interactions between the lead halide core of the 2D perovskite and TMD is not significantly affected by the presence of non-polar organic species or Rb at the heterointerface. Therefore, for accurate calculations at a reasonable computational cost, the properties of various 2D perovskite/TMD heterojunctions can be analysed using the corresponding Rb-substituted inorganic structures. For a more accurate description of the bandgap in HS2, the hybrid HSE06 functional was employed which resulted in 1.12 eV. Interestingly, the heterostructure comprising $MoS_2$ layer rotated by 90° and stacked on top of the 2D perovskite shows comparable electronic properties as demonstrated in Supporting Information S6. The formation enthalpies of the three heterostructures (Supporting Information S7) are negative, indicating thermodynamic stability. Next, we look at the bandstructure at the heterointerface of $BA_2PbBr_4$ with $WS_2$ (Supporting Information, S8) which demonstrates a nature comparable to the above-mentioned $MoS_2$ based system. For the all-inorganic $Rb_2PbBr_4/WS_2$ structure, DFT-GGA calculated bandgap is 1.04 eV and that obtained via HSE06 is 1.44 eV, close to the desirable value predicted by Shockley-Queisser theory for optimum photovoltaic efficiency.

The sizeable band gaps at the 2D heterointerface of $BA_2PbBr_4$ with monolayer $MoS_2$ or $WS_2$ allows the possibility of photoexcitation across a broad range of the electromagnetic spectrum including (i) visible (intralayer in $MoS_2$ or $WS_2$), (ii) near-UV (intralayer in $BA_2PbBr_4$) and (iii) near-IR (hetero-interface) as shown in Figure 2b. Moreover, the spatially separated interlayer excitons (shown in grey colour in Figure 2b) could be possible in these heterostructures in addition to intralayer excitons (black) localised in the same material. Sub-bandgap photodetection enabled by such interlayer excitons across the interfacial gap has been shown for the case of $MoS_2/WSe_2$[32] and $WSe_2/ReS_2$[33] based TMD heterostructures. Further, near-IR interlayer exciton emission involving electrons/holes at **K** valleys of both materials has also been reported for $MoS_2/WSe_2$ heterojunctions.[34] These observations suggest that $BA_2PbBr_4/MS_2$ heterostructures considered here can enhance the versatility of 2D perovskites by enabling interlayer emission or optical absorption in the visible and near-infrared energies, below the bandgap of the 2D perovskite.

Considering the bandgap character, for both $MS_2$-based heterostructures, the bandgap is indirect in nature, with the valence band maximum (VBM) at **Γ** and conduction band minimum (CBM) located along the high symmetry line **Γ–X**. The *k* point at which CBM occurs is related with the folding of the hexagonal Brillouin zone (BZ) of the TMD onto the orthorhombic BZ of the heterostructure. By unfolding the electronic states of the heterostructure into the hexagonal irreducible Brillouin zone (IBZ) of TMD, the point at CBM (**K′** along **Γ−X**) in the rectangular supercell BZ corresponds to the **K** point of the hexagonal IBZ of TMDs (Figure 2d). Further, the bandstructures unfolded onto the hexagonal IBZ of TMD as well as the orthorhombic IBZ of $BA_2PbBr_4$ can be found in the Supporting Information, S9 and show good resemblance with the pristine materials as can be seen from the overlaid picture. Therefore, the weak van der Waals interaction preserves these intrinsic characteristics of the constituents in the bandstructure of the hetero-material system. Moreover, the heterostructure comprising $MoSe_2$ and BA$_2$PbBr$_4$ (Supporting Information S8) exhibits features like the sulphide-based structures owing to similar band alignments.

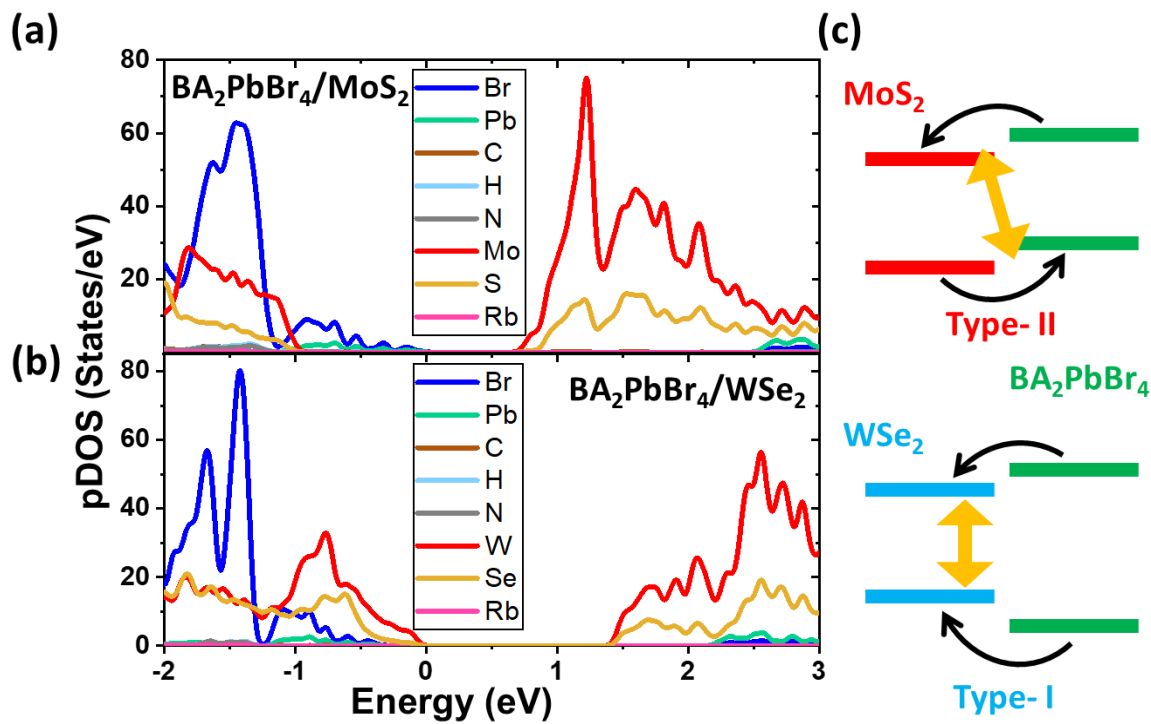

Figure 3: Atom projected density of states of (a) BA$_2$PbBr$_4$/MoS$_2$ (type-II, indirect) and (b) BA$_2$PbBr$_4$/WSe$_2$ (type-I, direct) heterostructures showing the contributions to energy states around the Fermi level. In both cases, the contribution of Rb from the corresponding all-inorganic structures is also included. Based on the nature of band edges in the two structures, the schematic in (c) illustrates the type of band alignment in both cases.

In contrast to the heterostructures discussed above, the bandstructure of selenide-based heterostructure BA$_2$PbBr$_4$/WSe$_2$ in Figure 2c demonstrates a direct bandgap at K' along Γ-X of the BZ and the DFT-GGA calculated ~~band~~ gap is around 1.55 eV, close to that of pristine WSe$_2$. The energy difference between the direct (K') and indirect (S-Y) gaps is low (<25 meV), hence the optical transitions across the direct bandgap will be more favourable. The DFT-GGA calculated band gap is around 1.55 eV, close to that of pristine WSe$_2$. The bands of the all-inorganic Rb$_2$PbBr$_4$/WSe$_2$ heterojunction (shown in Supporting Information, S6) depict similar characteristics in terms of the nature and gap value, thereby reinforcing our methodology on Rb-substitution of the BA group. The HSE06 formalism yields a higher bandgap of 2.02 eV. Further, from Figure S10 in Supporting Information, the unfolded bands of Rb$_2$PbBr$_4$/WSe$_2$ also show good correlation with that of the pristine WSe$_2$ and Rb$_2$PbBr$_4$. The bandstructures of Rb-substituted perovskites with MoS$_2$ and WSe$_2$ in presence of spin-orbit coupling (SOC) are shown in S11 of Supporting Information.

To understand the atom-wise contributions in the electronic bandstructure of 2D perovskite/TMD heterostructure, we examine the projected density of states calculated using gaussian smearing with a low smearing width of 0.05 eV. For the BA$_2$PbBr$_4$/MoS$_2$ heterostructure, the electronic states at the conduction band edge are made up of Mo and S (Figure 3a). The $4d_{z^2}$ orbitals of Mo give rise to energy levels at the CBM of the heterostructure just as in the case of monolayer MoS$_2$ while the Br $4p_{x,y,z}$ and Pb $5s$ orbitals of BA$_2$PbBr$_4$ constitute the VBM of the heterostructure. We observe similar electronic states in BA$_2$PbBr$_4$/WS$_2$ and BA$_2$PbBr$_4$/MoSe$_2$ (Supporting Information, S8), with the respective $4d_{z^2}$ and $5d_{z^2}$ of Mo and W dominating the conduction bands and $4p$ of Br mainly occupying the valence bands. Interestingly, in BA$_2$PbBr$_4$/MoSe$_2$, the valence band offset is quite low (0.09 eV), such systems can show a transition from type-II to type-I with external factors like strain. For all the Rb-substituted structures, the states from Rb are located deeper in the valence and conduction bands, away from the Fermi level. Therefore, we can conclude that a type-II band alignment is obtained for heterostructures of MoS$_2$, WS$_2$ and MoSe$_2$ with 2D A$_2$PbBr$_4$ (where A=butylammonium, BA or Rb) perovskite in which the valence and conduction band edges are contributed by the 2D perovskite and TMD, respectively.

The large exciton binding energies, around 300 meV for BA$_2$PbBr$_4$,[8] is a major limitation for photovoltaics using few-layer 2D perovskites since it can lead to a fast recombination of photogenerated charge carriers that could otherwise drive a photocurrent in the circuit. This has led to low power conversion efficiency values for the 2D perovskite solar cells as well as poor light detection.[35] However, the type-II band alignment of the heterostructures of MoS$_2$, WS$_2$ and MoSe$_2$ with 2D A$_2$PbBr$_4$ perovskite allows the formation of spatially separated interlayer excitons. The photogenerated electrons could reside in the sulphide based TMD while holes move to BA$_2$PbBr$_4$ in the case of interlayer excitons. For example, the sizeable conduction band offset (CBO) in BA$_2$PbBr$_4$/MoS$_2$ heterostructure of ~1.86 eV and valence band offset (VBO) of ~ 0.97 eV are larger than the intralayer exciton binding energy, which could promote the separation of these photocarriers across the heterojunction at low electric fields to generate large photocurrents.[15] The dissociation rate of interlayer excitons has been shown to be significantly higher than intralayer excitons for TMD based type-II interfaces due to their reduced binding energies.[36] The reduced interlayer bandgap due to the type-II band alignment and presence of high density of states at the valence and conduction band edges should enhance the absorption of photons in the near IR and visible regions of the electromagnetic spectrum in comparison to the individual materials. Hence, heterostructures of MoS$_2$, WS$_2$ and MoSe$_2$ with 2D A$_2$PbBr$_4$ perovskite can enable an improved excitonic solar cell performance as well as enhanced photodetection.

In contrast, for the BA$_2$PbBr$_4$/WSe$_2$ heterojunction, the conduction and valence band edges are both made up of the $5d_{x^2-y^2}$, $5d_{xy}$ orbitals of W and $4p_x$, $4p_y$ orbitals of Se (Figure 3b). Since both band edges are from the same material monolayer, the electronic structure demonstrates a type-I band alignment, and the band offsets (CBO ~ 0.79 eV and VBO ~ 0.34 eV) are small as shown in Figure 3b. This type-I configuration drives a unidirectional transport of the photo-excited electrons and holes in BA$_2$PbBr$_4$ to the WSe$_2$ layer. The accumulation of electrons and holes in WSe$_2$ could promote enhanced radiative recombination of these electron-hole pairs leading to light emission. A recent experimental study on *n*=4 2D perovskite/TMD heterostructure shows a 150-fold enhancement of PL emission and 25-times increase in optical absorption owing to the type-I band alignment at the hetero-interface.[37] Therefore, in our case, the direct bandgap values could enable carrier recombination-based devices in the near-infrared/visible, similar to that observed in monolayer ReS$_2$/MoS$_2$ type-I heterojunction.[38]

The electrical transport in 2D perovskite/TMD heterostructure will be strongly influenced by the effective mass (m*) of charge carriers at the interface which can be obtained from the bandstructures in Figure 2 using $m^* = \hbar^2 / \frac{\partial^2 E}{\partial k^2}$, where $\hbar$ is reduced Planck's constant, $E$ is energy, and $k$ is wave vector. For BA$_2$PbBr$_4$/MoS$_2$ heterostructure, the effective mass of holes (m$_h$) is found to be 0.29m$_0$ by using the parabolic approximation at the VBM edge. For electrons, m$_e$ is found to be 0.52m$_0$ at the conduction band minimum. For comparison, the effective masses of electrons and holes in the monolayer BA$_2$PbBr$_4$ are 0.22m$_0$ and 0.30m$_0$, respectively.[39] For monolayer MoS$_2$, the effective mass of electrons is 0.51m$_0$ and for holes is 0.60m$_0$. The effective masses of charge carriers in all 2D perovskite/TMD heterostructure are comparable to the individual materials, indicating that the carrier transport will not be significantly degraded upon the formation of vdW heterostructures.

We also explore the heterostructures based on BA$_2$PbI$_4$ and BA$_2$PbCl$_4$ which are members of the homologous series of $n$ = 1 2D perovskites. For monolayer BA$_2$PbI$_4$, our calculations yield a HSE06 band gap of 2.91 eV similar to other reports whereas absorption measurements show 2.38 eV.[8,25] The HSE06 gap of BA$_2$PbCl$_4$ is 3.92 eV while experimentally it has been found to be around 3.6 eV.[1] The heterostructure of BA$_2$PbI$_4$ with WSe$_2$ shows a type-II band alignment with an interlayer bandgap of 1.16 eV, different from the BA$_2$PbBr$_4$/WSe$_2$ structure. The CBO and VBO of BA$_2$PbI$_4$/WSe$_2$ are lower than the BA$_2$PbBr$_4$/MoS$_2$ heterostructure described earlier. A recent experimental study on (iso-BA)$_2$PbI$_4$/WSe$_2$ heterojunction shows the presence of interlayer excitons at around 1.6 eV, which is lower than the bandgaps of monolayer WSe$_2$ and (iso-BA)$_2$PbI$_4$. This confirms the type-II nature at the interface,[20] thereby independently validating these results.

For the BA$_2$PbI$_4$/MoS$_2$ heterojunction, the conduction and valence bands in DFT-GGA results nearly overlap along $\mathbf{\Gamma-X}$ of the BZ, leading to a transition from type-II to type-III alignment. On the other hand, HSE predicts a low bandgap value of 0.18 eV. This crossover of the bands could be due to the larger inherent strain (~ 4.1 %) inevitable for good lattice matching in the supercell. However, such a broken band alignment could be useful for tunnelling-based devices and the energy overlap for tunnelling could be tuned via a transverse electric field (gate) or external strain. In the above-mentioned experimental report also, no interlayer excitons are seen for the case of (iso-BA)$_2$PbI$_4$/MoS$_2$.[20] Further, the heterostructure of BA$_2$PbCl$_4$ with MoS$_2$ also depicts a type-II band alignment with sizeable band offsets and a gap of 1.29 eV. In both these heterostructures, the CBM and VBM are composed of MoS$_2$ and the 2D perovskite, respectively. The bandstructures and projected density of states plots for these heterostructures are provided in S12 of Supporting Information. In conjunction with our calculations, recent experimental evidence of interlayer excitons in 2D perovskite/monolayer TMD hetero systems indicate the strong interlayer electronic coupling mediated by charge transfer across the interface of these materials.

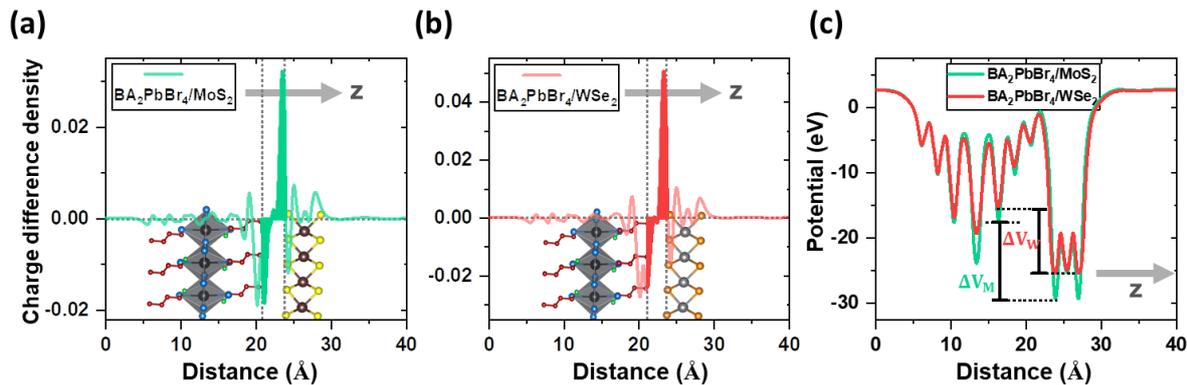

Figure 4: Planar averaged charge density difference profile along the direction normal to the interface (z) for (a) BA$_2$PbBr$_4$/MoS$_2$ and (b) BA$_2$PbBr$_4$/WSe$_2$ heterostructures. (c) Electrostatic potential profile along *z*- direction for both structures where the average potential difference across the interface for both systems ($\Delta V_W$, $\Delta V_M$) are marked.

To understand the extent of charge transfer through the van der Waals gap, we next examine the planar-averaged charge density difference along the direction perpendicular to the interface ($z$). Figure 4a shows a clear separation of charges across the interface in BA$_2$PbBr$_4$/MoS$_2$ structure as expected from the type-II alignment. The positive charge difference indicates electron accumulation at the MoS$_2$ edge and holes at BA$_2$PbBr$_4$ edge. While the BA$_2$PbBr$_4$/WSe$_2$ in Figure 4b shows a build-up of charges at the junction owing to the straddling gap. The average potential profiles along the $z$- axis for both structures are shown in Figure 4c. The peaks corresponding to the atoms of the TMD and 2D perovskite are seen.

From the analysis of the electronic structure, band alignments and interfacial charge transfer of various 2D perovskite/TMD heterostructures, different types of band alignments (type I, II and III) are feasible. This highlights the need for a first principles-based selection of individual constituents of the hetero-stack to guide specific experimental realizations. BA$_2$PbBr$_4$/WSe$_2$ heterostructure demonstrates a type-I alignment while BA$_2$PbI$_4$/MoS$_2$ is of type-III. The multiple possibilities of obtaining type-II alignment are provided by combining BA$_2$PbBr with MoS$_2$, WS$_2$ and MoSe$_2$ as well as BA$_2$PbI$_4$/WSe$_2$ and BA$_2$PbCl$_4$/MoS$_2$ structures, wherein the tunability in interface bandgaps is an important feature. Section S13 in Supporting Information lists the calculated carrier effective masses at band edges and the conduction and valence band offsets which are key properties that govern optoelectronic applications.

**Table 1:** Compilation of the electronic bandstructure properties (bandgaps in eV, nature, and alignment type of 2D perovskite/TMD heterostructures). HSE06 computed values are shown in parentheses.

|  | BA$_2$PbBr$_4$ | Rb$_2$PbBr$_4$ | BA$_2$PbI$_4$ | Rb$_2$PbI$_4$ | BA$_2$PbCl$_4$ | Rb$_2$PbCl$_4$ |
|---|---|---|---|---|---|---|
| MoS$_2$ | 0.77 | 0.73 (1.12) | 0.03 | 0.0 (0.18) | 1.29 | 1.21 (1.72) |
| WSe$_2$ | 1.55, direct | 1.52 (2.02), direct | 1.19 | 1.09 (1.49) | | |
| WS$_2$ | 1.12 | 1.04 (1.44) | | | | |
| MoSe$_2$ | 1.40 | 1.27 (1.67) | | | | |

| Type-I | Type-II | Type-III |
|---|---|---|

## Effect of strain:

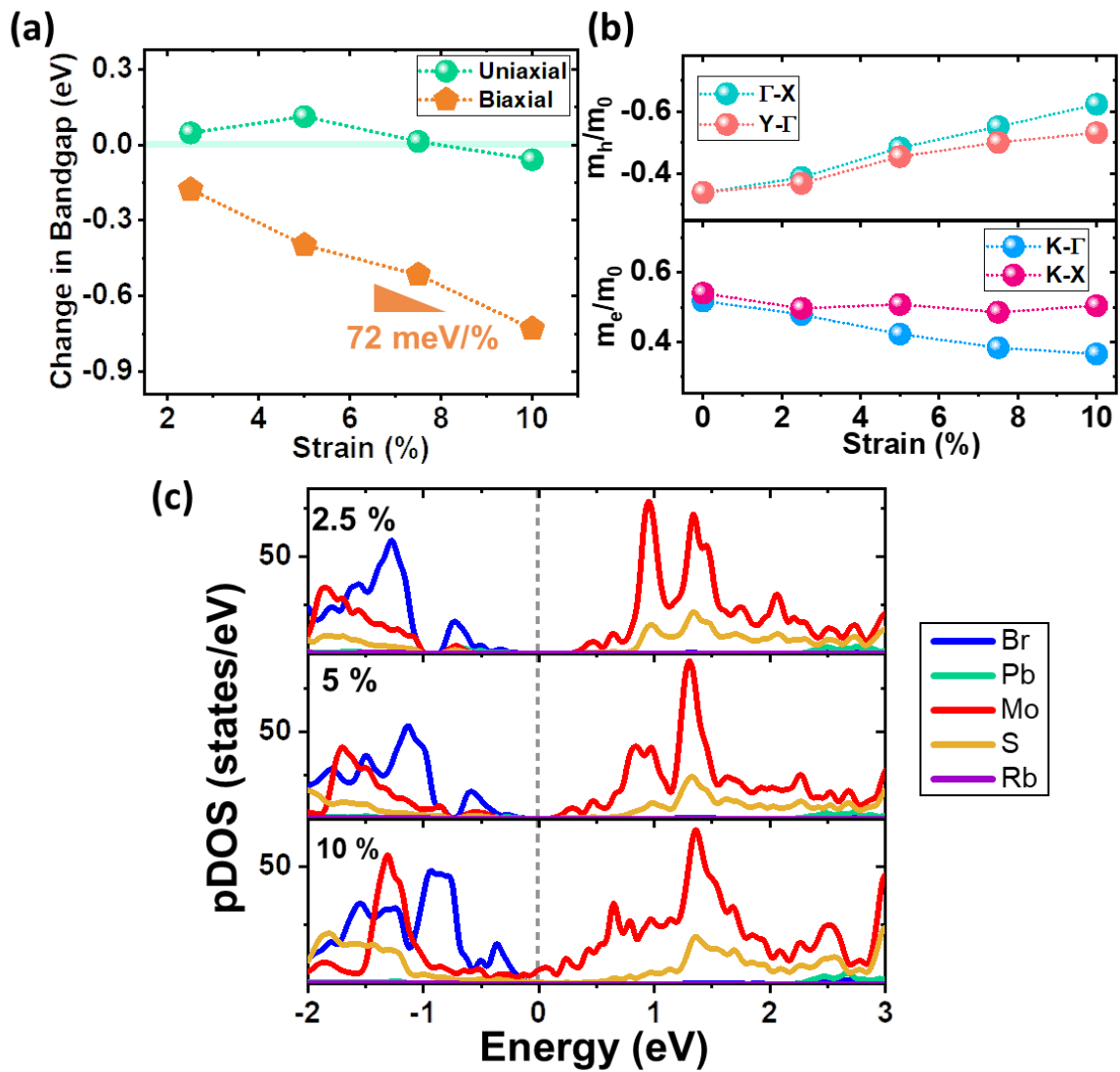

Figure 5: Effect of strain. (a) Dependence of uniaxial and biaxial strain on the bandgap of $Rb_2PbBr_4/MoS_2$ heterostructure. (b) Variation in the effective masses of electrons and holes with uniaxial strain. (c) Atom projected density of states showing the evolution of element-wise contributions to energy levels at conduction and valence band edges with the marked biaxial strain. The semiconductor-to-metal crossover at large biaxial strain can be attributed to the $d_{z^2}$ orbitals of $MoS_2$.

Next-generation applications in flexible solar cells and wearable (photo)sensors require materials which exhibit stable optoelectronic attributes under repeated strain and relaxation cycles. In this context, 2D perovskites have been shown to accommodate lattice strain without drastic structural deformation due to the bulky organic layers that provide enhanced stability to the layered 2D phase which is not present in the 3D bulk perovskites.[39] For 2D perovskites strained via tensile lattice expansion or laser-induced thermal stress, a monotonic decrease in bandgap has been experimentally observed[29,39] which is restored upon withdrawal of the external stress. Moreover, monolayer TMDs can also withstand considerable lattice strain owing to their high effective Young's modulus.[28]

As an illustrative example, here we employ uniaxial and biaxial tensile strain on the all inorganic, $Rb_2PbBr_4$/$MoS_2$ heterojunction to understand its effects on the electronic structure and associated properties. We apply uniaxial tensile strain from 0 to 10 % applied along the $y$-direction and the ions are allowed to relax in the $x$-direction which is expected to redistribute the effect of strain on the structure. In the $Rb_2PbBr_4$/$MoS_2$ heterostructure, with increasing uniaxial strain, the bandgap shows a slight increase up to 5% strain and thereafter, it decreases at a rate of 34 meV/% strain. The experimentally observed variation of bandgap is 13.3 meV/% in $n = 5$ 2D perovskite (strain up to 1.18%)[40] which is smaller than the 45 meV/% seen in single layer $MoS_2$.[41] The change in bandgap values and evolution of bandstructures with uniaxial tensile strain are shown in Figure 4a and S14 of Supporting Information, respectively. In terms of electronic properties, the effective masses of electrons show a slightly decreasing trend at higher uniaxial tensile strain and that of holes shows an increase (Figure 5b), but are comparable to the unstrained structure, confirming that the carrier transport properties remain mostly unchanged. For uniaxial strains, the CBM contributed by $MoS_2$ shows a shift away from the **K'** point of the BZ towards **X** while the VBM remains at **Γ**. The tensile stretching of the supercell increases the average equatorial (Pb-Br-Pb) bond angles and elongates the average Pb-Br bonds, leading to distortions in the lead-halide octahedra which affects the electronic bandstructure.

In comparison to one-dimensional lattice stretching, the lowering of the interface bandgap is more pronounced when biaxial strain is applied by stretching the unit cell along both *x* and *y* directions. Particularly, at a large biaxial strain of 10 % the conduction and valence bands cross-over, resulting in a semiconductor-to-metal transition (type-II to -III) in $Rb_2PbBr_4/MoS_2$ heterostructure. In monolayer $MoS_2$, with the application of tensile strains, the valence band maximum at **Γ** shifts upward relative to that at **K** and at large biaxial stretching a significant decrease in bandgap leading to semiconductor-to-metal transition is predicted by first-principles calculations.[42] In the heterostructure, the sharp decrease in bandgap and ultimately the crossing-over can be attributed to $MoS_2$ from the projected density of states of the strained structures as shown in Figure 5c. Further, the orbital contributions reveal that the $d_{z^2}$ orbitals of Mo from valence and conduction bands overlap at 10 % biaxial strain. The bandstructures of the biaxially strained structures are shown in S13 of Supporting Information. From the above analysis, we can conclude that under externally applied strains, dynamic reorientation of the lead-halide octahedra as well as the linked organic chains could take place in the 2D perovskite lattice.[43] Furthermore, the variation in electronic properties with strain is seen to be consistent for various TMDs,[44] hence the response of the various other 2D perovskite/TMD heterostructures towards tensile strain is expected to be fairly similar to the trends described here.

**Device and growth aspects:**

TMDs have been employed as selective charge carrier transport layers to improve the efficiencies of bulk perovskite solar cells, specifically as electron transport layer (ETL) in solar cells with $MAPbI_3$ absorber layer.[45] In the type II $A_2PbBr/MS_2$ heterojunctions considered here, the TMD can enable the transport of photogenerated electrons away from the 2D perovskite and at the same time block the transport of holes owing to the significant band offsets thereby reducing charge recombination. Therefore, the protective layer of TMDs on top of the 2D perovskite can be an effective ETL for improved power conversion efficiencies.

In bare 2D perovskite layers, external strain could result in the loss of organic species and greater reactivity towards moisture which can impair the hybrid 2D framework and cause structural damage as in the case of 3D perovskites.[29] Therefore, in our heterostructures, the TMD as a capping layer also functions as an impermeable protection against oxygen and moisture which has been shown to degrade the properties of strained 2D perovskites at a faster rate.[46] Hence, the 2D TMD in heterojunction with 2D perovskite can impart better stability to the latter for flexible device applications.

A commonly encountered experimental challenge while fabricating heterostructures with 3D perovskites like $MAPbI_3$ is the possibility of multiple edge terminations at the interface- either PbI or MAI. DFT calculations have shown that in comparison to the PbI-terminated surfaces, the MAI terminations can have CBM and VBM up to 1 eV lower in energy, thereby resulting in experimental inconsistencies.[47] However, in case of 2D perovskites, the van der Waals interaction holding the adjacent layers and absence of dangling bonds could envisage better interfacial properties. Recently, vapor phase growth of 2D perovskite layers on TMDs has demonstrated a good promise for photodetection because of the strong interlayer coupling.[48] Chemical vapor deposition (CVD) based growth of TMDs has been successful in yielding large area films with excellent electrical and optoelectronic properties. Along with that, CVD and vapor phase deposition methods have been employed for lateral and vertical TMD heterostructures. Therefore, other solution- and/or CVD-based processes at moderate temperatures need to be developed to enable the large area growth of 2D perovskite layers on TMD templates for photovoltaic applications.

To summarise, we have systematically explored the properties of the heterojunctions comprising 2D perovskite and TMDs for optoelectronic applications using first principles calculations. The presence of an interlayer bandgap at the interface of heterostructures of $BA_2PbBr_4$ with $MoS_2$, $MoSe_2$ and $WS_2$ enables optical absorption at energies much lower than its large intrinsic bandgap of ~ 3 eV. The type-II band alignment in these heterostructures could lead to interlayer separation of photogenerated carriers with TMD as an efficient electron transport layer and help to overcome the large intralayer exciton binding energy. The heterostructure also shows a good tolerance towards externally applied tensile strain with low modulation of the bandgap with unidirectional strain, essential for flexible technologies. Additionally, the presence of a type-I bandgap at the $BA_2PbBr4/WSe_2$ heterojunction makes it ideal for light emission applications in the visible range. Further, the type-III alignment in $BA_2PbI_4/MoS_2$ structure could open up applications for 2D perovskites for tunnelling devices. The capping of 2D perovskite with TMD monolayer can render enhanced air stability and device longevity. Furthermore, development of van der Waals epitaxy or other growth techniques could enable the creation of large area devices based on 2D perovskite/TMD heterostructures for next-generation sensors for internet of things-based applications. Understanding the carrier transport properties across the different heterostructures could shed more light on the device-level application of these heterostructures.

**Methods:**

In all our atomistic simulations implemented using Vienna Ab initio Simulation Package (VASP), the generalized gradient approximation of Perdew-Burke-Ernzehof form was used to describe the exchange correlation[49,50] and projector augmented wave (PAW) pseudopotentials were employed to account for the frozen core-valence electron interactions. To introduce van der Waals interaction between the two materials, the dispersion corrected DFT-D3 method was utilized.[51] For structural optimization, a plane wave cut-off of 400 eV and Γ-centred k-points grid of 5×5×1 were used. The self-consistent calculations used a higher energy cut-off of 500 eV and denser k-points grid of 9×9×1. The optimizations were performed until the energy convergence limit of 0.1 meV was reached and the Hellman-Feynman forces on the structures were less than 0.01 eV/Å. Sufficient vacuum of at least 25 Å was incorporated in all supercells along the transverse direction to avoid spurious interactions between adjacent periodic images. Since the bandgaps are typically underestimated by the GGA-PBE formalism, hybrid calculations using HSE06 scheme were also used for accurate prediction of bandgaps.[52] Maximally localised Wannier wavefunctions using Wannier90 were employed for HSE bandstructure calculations.[53] VASPKIT tool was used for post processing of the calculated data.[54] VESTA was used to visualize the crystal structures.

**Acknowledgements:**

The authors gratefully acknowledge computational support from the Australian National Computing Infrastructure (NCI) and Pawsey supercomputing facility for high performance computing. Y.Y. and N.V.M. acknowledge the support from the Australian Research Council (CE170100039). S.L acknowledges support from the Department of Science and Technology, Govt. of India through its SwarnaJayanti fellowship scheme (Grant No. DST/SJF/ETA-01/2016-17)

**Competing Interests:**

The authors declare no competing interests.

# Supporting Information

**Near-Infrared and Visible-range Optoelectronics in 2D Hybrid Perovskite/Transition Metal Dichalcogenide Heterostructures**

Abin Varghese,[1,2,3,4] Yuefeng Yin,[1,4] *, Mingchao Wang,[1,5] Saurabh Lodha,[2] Nikhil V. Medhekar[1,4]*

Email: yuefeng.yin@monash.edu, nikhil.medhekar@monash.edu

[1] Department of Materials Science and Engineering, Monash University, Clayton, Victoria 3800, Australia

[2] Department of Electrical Engineering, Indian Institute of Technology Bombay, Mumbai 400076, India

[3] IITB-Monash Research Academy, IIT Bombay, Mumbai 400076, India

[4] ARC Centre of Excellence in Future Low-Energy Electronics Technologies, Monash University, Clayton, Victoria 3800, Australia

[5] Centre for Theoretical and Computational Molecular Science, Australian Institute for Bioengineering and Nanotechnology, The University of Queensland, St Lucia, QLD 4072, Australia

## S1. Lattice parameters of TMDs and 2D perovskites

Lattice parameters of TMDs and 2D perovskites calculated using GGA including the van der Waals correction (DFT-D3), used in this work.

| Material | a (Å) | b (Å) | BZ |
|---|---|---|---|
| $MoS_2$ | 3.164 | 3.164 | Hexagonal |
| $WS_2$ | 3.169 | 3.169 | Hexagonal |
| $MoSe_2$ | 3.294 | 3.294 | Hexagonal |
| $WSe_2$ | 3.288 | 3.288 | Hexagonal |
| $BA_2PbI_4$ | 8.581 | 8.384 | Orthorhombic |
| $BA_2PbBr_4$ | 8.142 | 7.942 | Orthorhombic |
| $BA_2PbCl_4$ | 7.856 | 7.648 | Orthorhombic |

Table 1: Lattice parameters of TMDs and 2D perovskites using DFT-D3.

We also compare the lattice parameters and bandgap values using various van der Waals correction methods for the representative systems- $MoS_2$ for the monolayer TMDs and $BA_2PbBr_4$ for 2D perovskites. In addition to DFT-D3, for reference, we have employed the dispersion corrected DFT-D2 as well as 'opt' functional optB86vdW to account for the long-range van der Waals interactions, as shown below. It can be seen that DFT-D3 provides good correlation with experimental lattice parameters. However, the bandgap values are underestimated by all van der Waals functionals.

| $MoS_2$-Properties | DFT-D3 (Manuscript) | DFT-D2 | optB86-vdW | Expt. |
|---|---|---|---|---|
| Lattice Parameters (Å) | 3.163 | 3.189 | 3.163 | 3.15 |
| Bandgap (eV) | 1.72 | 1.62 | 1.73 | ~1.9 |

Table 2: Comparison of lattice parameters and bandgap of monolayer $MoS_2$ using various functionals that include van der Waals corrections.

| $BA_2PbBr_4$-Properties | DFT-D3 (Manuscript) | DFT-D2 | optB86-vdW | Expt. |
|---|---|---|---|---|
| Lattice Parameters (Å) | a= 8.142<br>b= 7.943<br>c= 26.233 | a= 8.046<br>b= 7.939<br>c= 25.939 | a= 8.124<br>b= 7.932<br>c= 26.218 | a= 8.253<br>b= 8.137<br>c= 27.402 |
| Bandgap (eV) | 2.50 | 2.39 | 2.42 | ~3 |

Table 3: Comparison of lattice parameters and bandgap of $BA_2PbBr_4$ using various functionals that include van der Waals corrections.

## S2. Bandstructure and density of states of 2D perovskites

Computed bandstructures of the homologous series of n=1 2D perovskites using the hybrid HSE06 functional incorporating van der Waals corrections (D3).

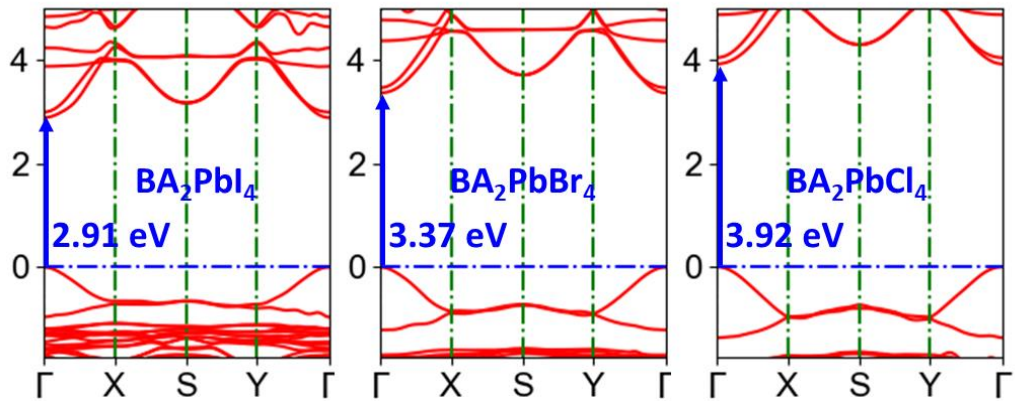

Comparison of bandgap values of n=1 2D perovskites computed used GGA and HSE06, both incorporating van der Waals correction (D3).

| 2D perovskite | GGA (eV) | HSE06 (eV) |
|---|---|---|
| $BA_2PbI_4$ | 2.08 | 2.91 |
| $BA_2PbBr_4$ | 2.50 | 3.37 |
| $BA_2PbCl_4$ | 2.91 | 3.92 |

Atom projected density of states of various 2D perovskites (D3) showing contributions of Pb and halogen in the electronic states at the band edges. The states due to the organic part (C, H, N) are deeper into the conduction and valence bands.

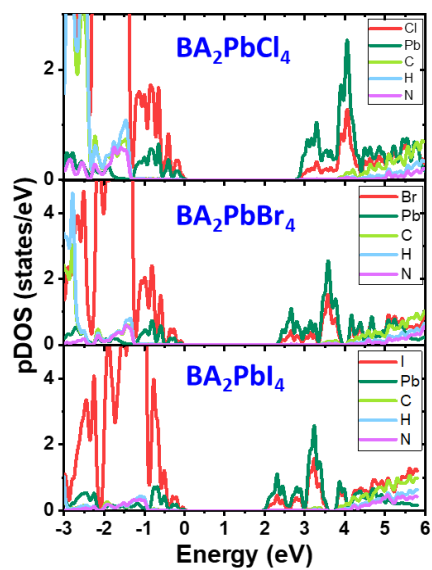

## S3. Bandgaps of monolayer TMDs

Bandgap values of monolayer TMDs computed by incorporating the dispersion corrected DFT-D3 in GGA as well as HSE06.

| TMD | GGA (eV) | HSE06 (eV) |
|---|---|---|
| $MoS_2$ | 1.72 | 2.23 |
| $WS_2$ | 1.84 | 2.35 |
| $MoSe_2$ | 1.51 | 1.98 |
| $WSe_2$ | 1.62 | 2.14 |

Comparison of the bands of monolayer $MoS_2$ in hexagon unit cell (a) and rectangular unit cell (b).

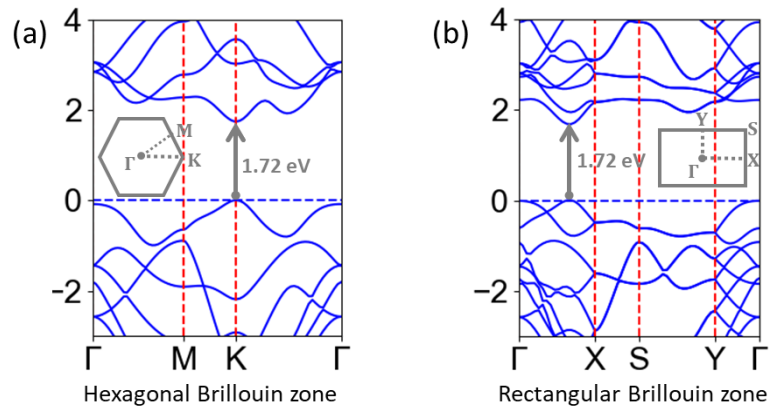

## S4. Strain in 2D perovskite/TMD heterostructures

In the 2D perovskite/TMD heterostructure, strain is split equally between the two materials. Table below shows the values of strain along the two in-plane directions (x,y) in different heterostructures considered in this work.

| 2D perovskite | TMD | Strain-x (%) | Strain-y (%) |
|---|---|---|---|
| $BA_2PbBr_4$ | $MoS_2$ | 1.42 | 1.72 |
| $BA_2PbBr_4$ | $WS_2$ | 1.35 | 1.83 |
| $BA_2PbBr_4$ | $MoSe_2$ | 0.57 | 3.74 |
| $BA_2PbBr_4$ | $WSe_2$ | 0.51 | 3.81 |
| $BA_2PbI_4$ | $MoS_2$ | 4.05 | 0.98 |
| $BA_2PbI_4$ | $WSe_2$ | 2.11 | 0.96 |
| $BA_2PbCl_4$ | $MoS_2$ | 3.60 | 3.48 |

## S5. Mixed 2D perovskite BARbPbBr$_4$ and its heterostructure

In addition to hybrid organic-inorganic and the fully-inorganic (Rb-substituted) 2D perovskites, we introduce a mixed structure BARbPbBr$_4$ represented below in which organic BA groups on one side of the 2D lead-halide octahedra are substituted by Rb. The computed bandstructure of this system is similar to BA$_2$PbBr$_4$ and Rb$_2$PbBr$_4$.

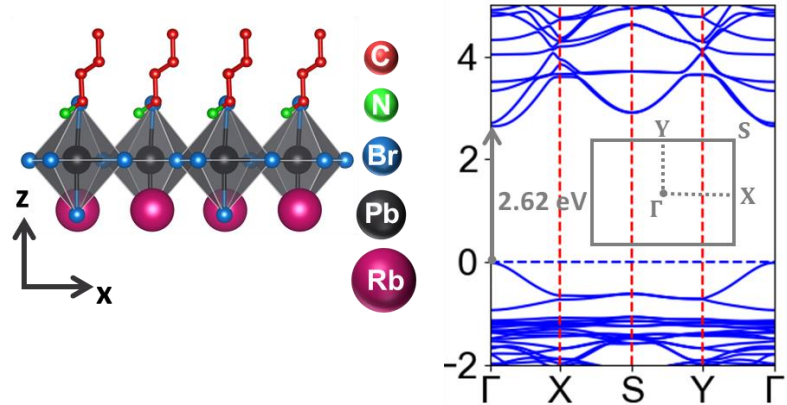

In the BARbPbBr$_4$/MoS$_2$ heterostructure (HS3) shown below, the calculated bandstructure exhibits features similar to HS1 and HS2.

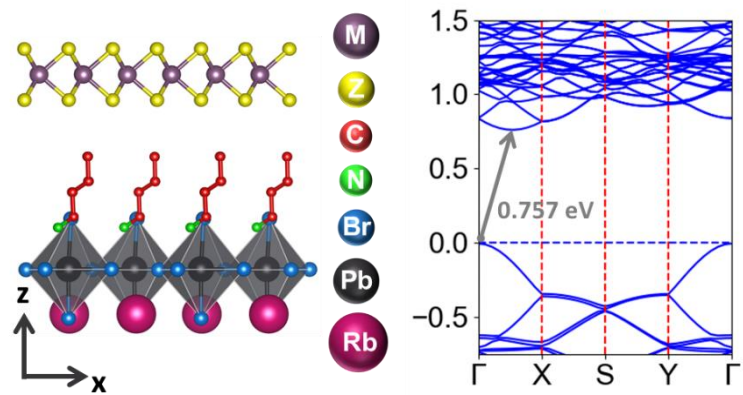

## S6. Bandstructures of all-inorganic Rb-substituted 2D perovskite heterostructures

Calculations of electronic properties of 2D perovskite/TMD heterojunctions in the manuscript have been performed using both the hybrid BA system as well as the corresponding Rb-substituted structure for the 2D perovskite. The bands of the Rb$_2$PbBr$_4$ constituted heterostructures with different TMDs are shown below.

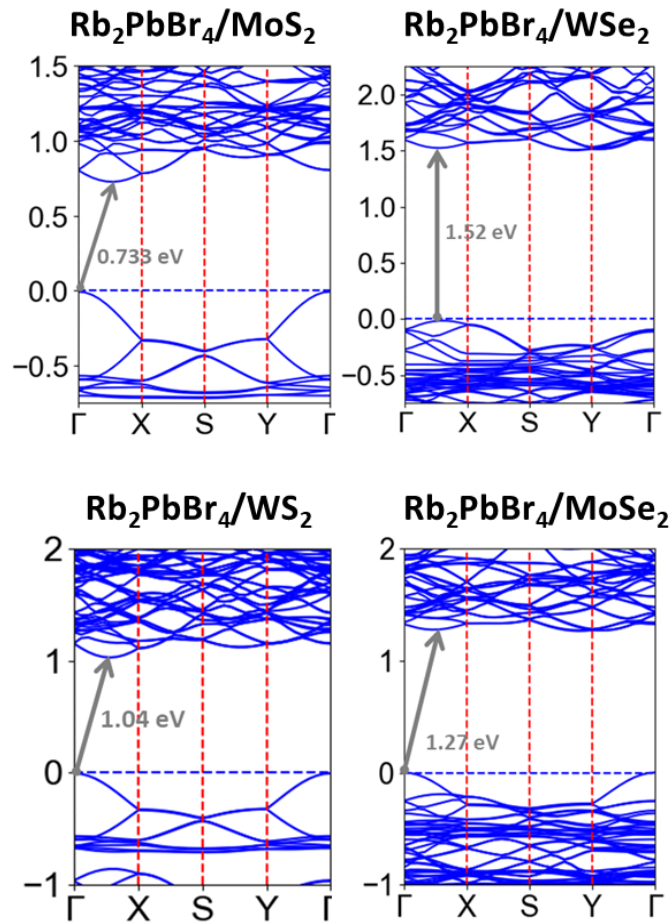

Additionally, the electronic bandstructure of the heterostructure in which the TMD layer on top of the 2D perovskite has been rotated by 90° is shown below. In the rotated MoS$_2$ stacked on Rb$_2$PbBr$_4$ (R90) system, the strain values are also reasonably low. The conduction band minimum occurs along Y → Γ due to the rotation of the TMD layer.

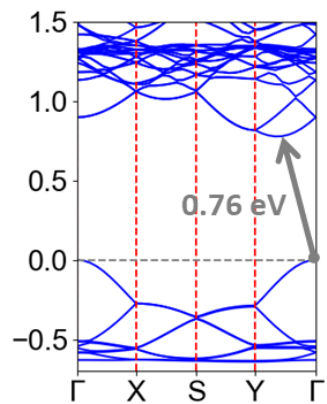

The electronic properties are compared in the table below:

| Properties | R90 (PBE) | 2×2 $Rb_2PbBr_4$/5×3 $MoS_2$ |
|---|---|---|
| Bandgap (eV) | 0.76 | 0.73 |
| $m_e^*$ ($m_0$) | 0.49 | 0.46 |
| $m_h^*$ ($m_0$) | 0.37 | 0.37 |

## S7. Formation enthalpies of heterostructures

We also calculated the formation enthalpies ($E_f$) for the three different structures (HS1, HS2 and HS3) to understand the thermodynamic stabilities using, $E_f = E_{HS} - E_P - E_{MoS_2}$, where $E_{HS}$, $E_P$ and $E_{MoS_2}$ are the total energies of the respective heterostructures, 2D perovskite and TMD components. The formation enthalpies per functional unit of the 2D perovskite are negative for each case (-0.45 eV for HS1, -0.03 eV for HS2, and -0.45 eV for HS3), indicating that the heterostructures are thermodynamically favourable.

## S8. Heterostructures of $BA_2PbBr_4$ with $WS_2$ and $MoSe_2$

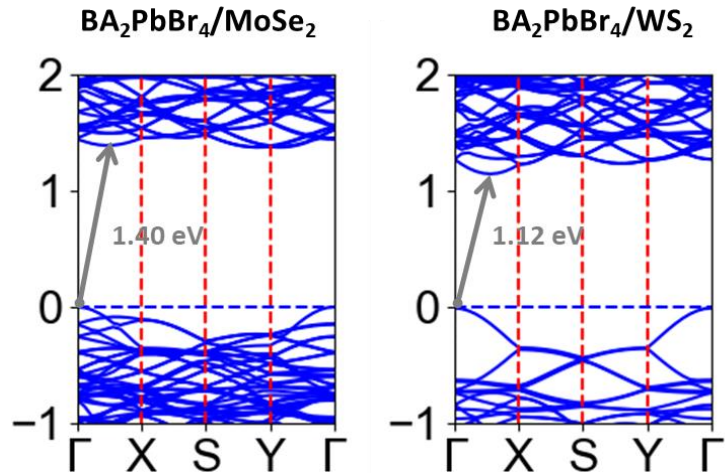

Atom projected density of states of the heterostructures:

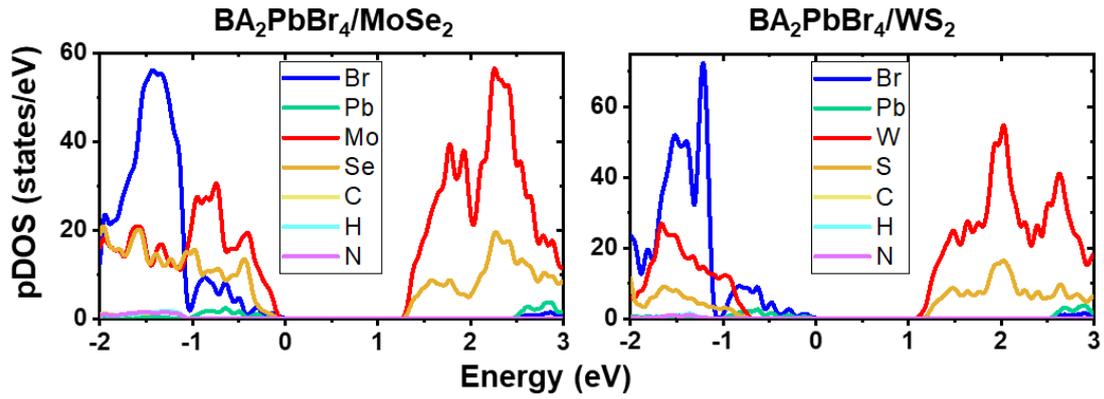

## S9. Unfolding of Rb$_2$PbBr$_4$/MoS$_2$ heterostructure bands

The bands of the heterostructure supercell have been projected onto the irreducible Brillouin zone of constituent materials for understanding distinct contributions using the *k*-projection method. Here, the scatter plot is the outcome of the projection and the solid lines represent the bandstructure of unstrained monolayer MoS$_2$ and Rb$_2$PbBr$_4$, in the respective figures. The solid lines are overlaid with the *k*-projection results as a guide for visualization.

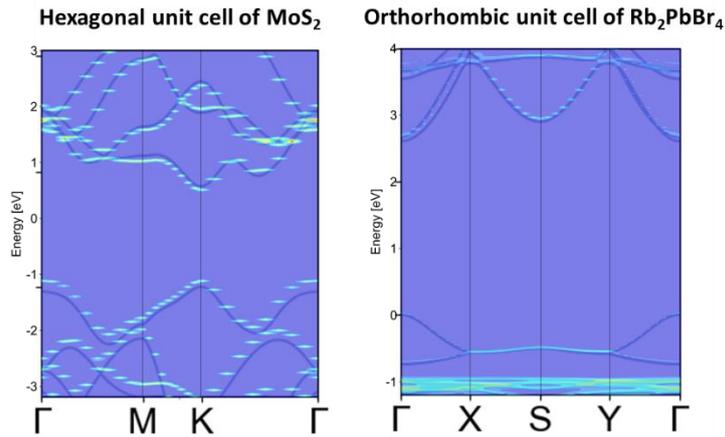

## S10. Unfolding of Rb$_2$PbBr$_4$/WSe$_2$ bands

The bands of the Rb$_2$PbBr$_4$/MoS$_2$ supercell have been projected onto the irreducible Brillouin zones of the individual materials. As described earlier, the solid lines are the bands of the pristine materials.

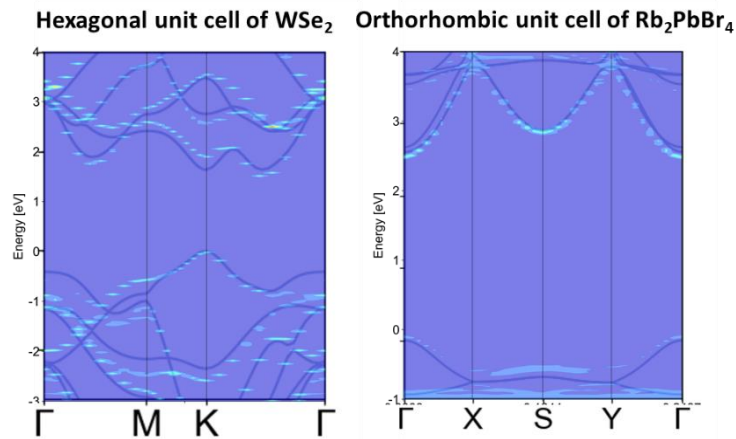

## S11: Effect of Spin-Orbit coupling (SOC) on band properties

To evaluate the effect of spin-orbit coupling in the electronic properties of 2D perovskite/TMD heterostructures, we have calculated the bandstructures of Rb$_2$PbBr$_4$/MoS$_2$ and Rb$_2$PbBr$_4$/WSe$_2$ with SOC turned on at the DFT-GGA level. In both cases, the band features around the Fermi level are preserved with respect to the non-SOC calculations. For Rb$_2$PbBr$_4$/MoS$_2$, the interface bandgap is 0.68 eV, slightly reduced in comparison to the non-SOC bandgap (0.73 eV). On the other hand, for Rb$_2$PbBr$_4$/WSe$_2$, the bandgap is lowered to 1.22 eV at **K** ( along **Γ-X**) of the Brillouin zone. The bandstructure plots are shown below.

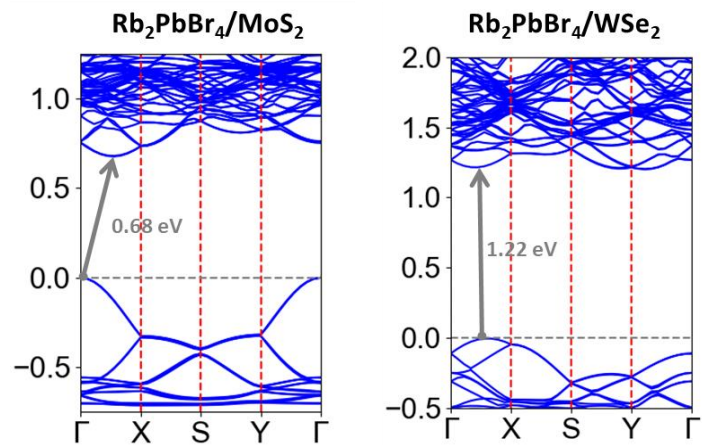

## S12. Bandstructures of TMDs with BA$_2$PbI$_4$ and BA$_2$PbCl$_4$

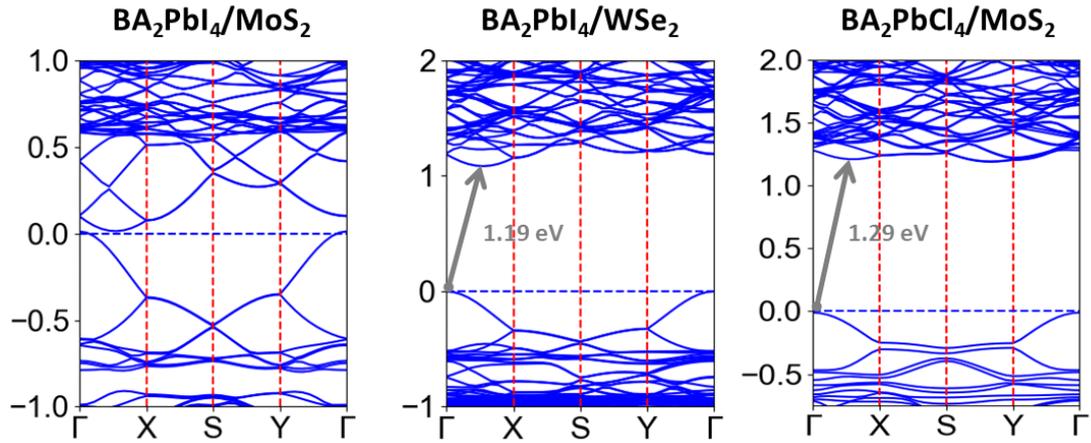

Density of states of the corresponding Rb-substituted heterostructures

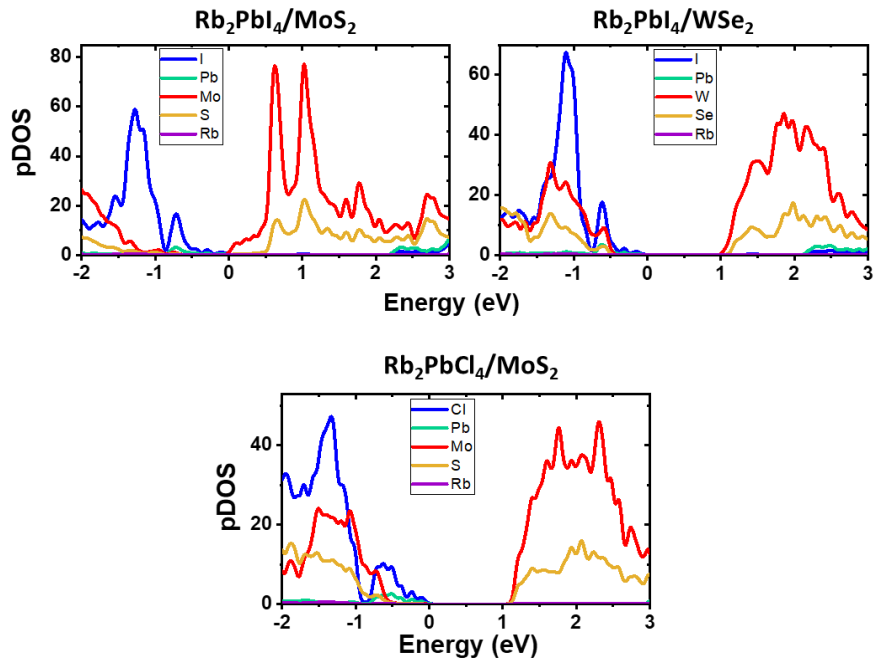

## S13. Summary of key calculated parameters of various 2D perovskite/TMD heterostructures

| 2D pvskt | TMD | $E_{G,GGA}$(eV) | $\Delta E_C$(eV) | $\Delta E_V$(eV) | $m_e^*$ | $m_h^*$ | Type |
|---|---|---|---|---|---|---|---|
| BA$_2$PbBr$_4$ | MoS$_2$ | 0.77 | 1.86 | 0.97 | 0.52 | 0.29 | II |
| BA$_2$PbBr$_4$ | WS$_2$ | 1.12 | 1.44 | 0.72 | 0.33 | 0.32 | II |
| BA$_2$PbBr$_4$ | MoSe$_2$ | 1.40 | 1.20 | 0.09 | 0.54 | 0.28 | II |
| BA$_2$PbBr$_4$ | WSe$_2$ | 1.55 | 0.79 | 0.34 | 0.53 | 0.54 | I, direct |
| Rb$_2$PbBr$_4$ | MoSeS (Janus) | 0.86 | 1.86 | 0.37 | 0.69 | 0.38 | II |
| **Rb₂PbBr₄** | **WSeS (Janus)** | 0.98 | 1.67 | 0.46 | 0.62 | 0.34 | II |
| **BA₂PbI₄** | **MoS₂** | 0.03 | - | - | 0.45 | 0.23 | II-III |
| **BA₂PbI₄** | **WSe₂** | 1.19 | 0.98 | 0.33 | 0.36 | 0.25 | II |
| **BA₂PbCl₄** | **MoS₂** | 1.21 | 1.84 | 0.45 | 0.68 | 0.45 | II |

S14. Evolution of bandstructure of Rb₂PbBr₄/MoS₂ with uniaxial strain

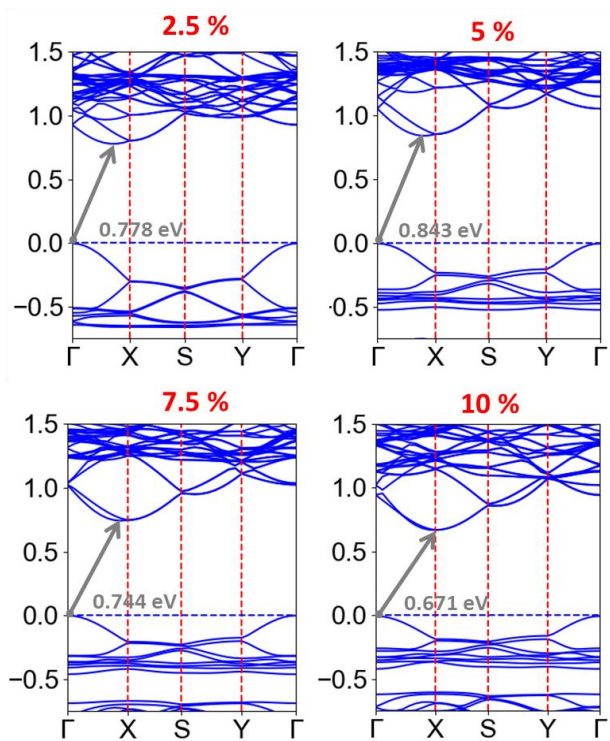

Evolution of bandstructure of $Rb_2PbBr_4/MoS_2$ with biaxial strain

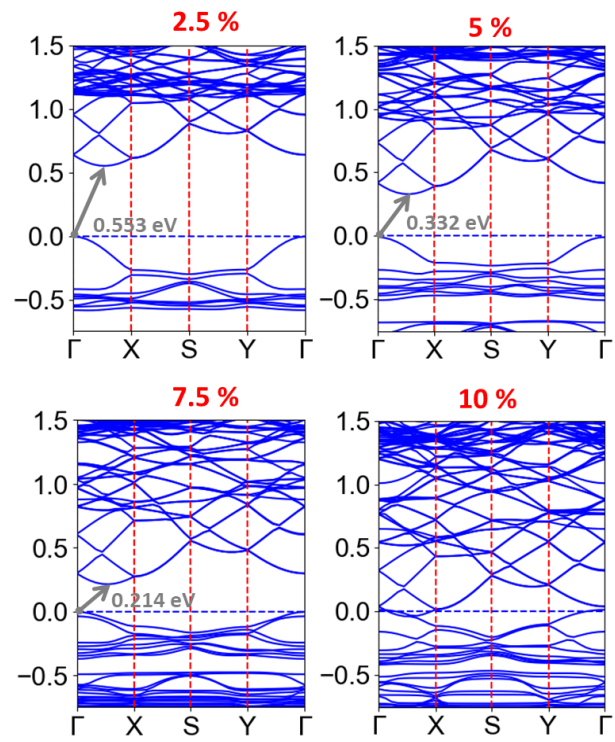